\documentclass[twoside,preprintnumbers,amsmath,amssymb,pacs]{revtex4}
\usepackage{epsfig}
\usepackage{graphicx}% Include figure files
\usepackage{dcolumn}% Align table columns on decimal point
\usepackage{bm}% bold math

\usepackage{amsthm} %%%%%

\newtheoremstyle{jvk-thm} % %%%%%
   {}{}{\itshape}{}{\bfseries}{.}{ }{} %%%%%
\newtheoremstyle{jvk-rem} % %%%%%
   {}{}{\upshape}{}{\bfseries}{.}{ }{} %%%%%

\theoremstyle{jvk-thm} %%%%%
\newtheorem{definition}{Definition}[section]

\theoremstyle{jvk-rem} %%%%%

\newtheorem{remarkth}[definition]{Remark}

\usepackage{pslatex}

\newcommand{\pslash}{p \! \! \! /}

\newcommand{\Nf}{N_{\!f}}
\newcommand{\NF}{N_{\!F}}
\newcommand{\MSbar}{\overline{\mbox{MS}}}
\newcommand{\oB}{\overline{B}}
\newcommand{\occ}{\overline{c}}
\newcommand{\oG}{\overline{G}}
\newcommand{\p}{\partial}
\newcommand{\NA}{N_{\!A}}

\begin{document}
\preprint{LTH-703}
\title{{\Large  Quantum properties of a non-Abelian gauge invariant action with a mass parameter}}

\author{M.A.L. Capri$^a$}
\email{marcio@dft.if.uerj.br}
\author{D. Dudal$^b$}\email{david.dudal@ugent.be}
\author{J.A. Gracey$^{c}$}
    \email{jag@amtp.liv.ac.uk}
\author{V.E.R. Lemes$^{a}$}\email{vitor@dft.if.uerj.br}
\author{R.F. Sobreiro$^a$}
 \email{sobreiro@dft.if.uerj.br}
\author{S.P. Sorella$^a$}
\email{sorella@uerj.br} \altaffiliation{Work supported by FAPERJ,
Funda{\c c}{\~a}o de Amparo {\`a} Pesquisa do Estado do Rio de
Janeiro, under the program {\it Cientista do Nosso Estado},
E-26/151.947/2004.}
\author{H. Verschelde$^b$}
 \email{henri.verschelde@ugent.be}
 \affiliation{\vskip 0.1cm $^a$ UERJ - Universidade do Estado do Rio de
Janeiro\\Rua S\~{a}o Francisco Xavier 524, 20550-013
Maracan\~{a}\\Rio de Janeiro, Brasil\\\\\vskip 0.1cm $^b$ Ghent
University
\\ Department of Mathematical
Physics and Astronomy \\ Krijgslaan 281-S9 \\ B-9000 Gent,
Belgium\\\\
\vskip 0.1cm $^c$ Theoretical Physics Division\\ Department of
Mathematical Sciences\\ University of Liverpool\\ P.O. Box 147,
Liverpool, L69 3BX, United Kingdom }

%\received{on 25 March, 2006}

\begin{abstract}
We continue the study of a local, gauge invariant Yang-Mills action
containing a mass parameter, which we constructed in a previous
paper starting from the nonlocal gauge invariant mass dimension two
operator $F_{\mu\nu} (D^2)^{-1} F_{\mu\nu}$. We return briefly to
the renormalizability of the model, which can be proven to all
orders of perturbation theory by embedding it in a more general
model with a larger symmetry content. We point out the existence of
a nilpotent BRST symmetry. Although our action contains extra
(anti)commuting tensor fields and coupling constants, we prove that
our model in the limit of vanishing mass is equivalent with ordinary
massless Yang-Mills theories. The full theory is renormalized
explicitly at two loops in the $\MSbar$ scheme and all the
renormalization group functions are presented. We end with some
comments on the potential relevance of this gauge model for the
issue of a dynamical gluon mass generation.
\end{abstract}

\maketitle

\setcounter{page}{1}

\section{Introduction.}
Yang-Mills gauge theories, with quantum chromodynamics (QCD)
modeling the strong interaction between elementary particles as one
of the key examples, are quite well understood at very high
energies. In this energy region, asymptotic freedom
\cite{Gross:1973id,Politzer:1973fx,Jones:1974mm,Caswell:1974gg} sets
in, which in turn ensures that the coupling constant $g^2$ is small
enough to make a perturbative expansion in powers of $g^2$ possible.
The elementary QCD excitations are the gluons and quarks.

Our current understanding of non-Abelian gauge theories is still
incomplete in the infrared region. At lower energies, the
interaction grows stronger, preventing the use of standard
perturbation theory to obtain relatively acceptable results.
Nonperturbative aspects of the theory come into play. The most
notable, yet to be rigourously proven nonperturbative phenomenon, is
the fact that the elementary gluon and quark excitations no longer
belong to the physical spectrum, being \emph{confined} into
colorless states such as glueballs, mesons and baryons.

A widely used strategy to parametrize certain nonperturbative
effects of the theory amounts to the introduction of so called
condensates, which are the expectation values of certain operators
in the vacuum. Furthermore, one can employ the operator product
expansion (OPE) (viz. short distance expansion) which can be applied
to \emph{local} operators, in order to relate the associated
condensates to nonperturbative power corrections which, in turn,
give additional information next to the perturbatively calculable
contributions.

As we are considering a gauge theory, these condensates should be
gauge invariant if they are to enter physical observables. This puts
rather strong restrictions on the possible condensates, the ones
with lowest dimensionality are the dimension three quark condensate
$\left\langle \overline{\psi}\psi\right\rangle$ and the dimension
four gluon condensate $\left\langle F_{\mu\nu}^2\right\rangle$.
There is a variety of methods to obtain estimates of these
condensates, such as the phenomological approach based on the SVZ
sum rules, \cite{Narison:2005hb} for a recent overview, the use of
lattice methods, \cite{DiGiacomo:1998nu}, as well as the use of
instanton calculus, \cite{Schafer:1996wv}.

Gauge condensates are necessarily nonperturbative in nature, as
gauge theories do not contain a mass term in the action due to the
requirement of gauge invariance. However, through nonperturbative
effects, a nontrivial value for e.g. $\left\langle
F_{\mu\nu}^2\right\rangle$ can arise.

In \cite{Lavelle:1988eg}, it was already argued that also gauge
variant condensates could influence gauge variant quantities such as
the gluon propagator. In particular, the dimension two gluon
condensate $\left\langle A_\mu^2\right\rangle$ has received much
attention in the Landau gauge
\cite{Gubarev:2000eu,Gubarev:2000nz,Kondo:2001nq,Verschelde:2001ia,Boucaud:2001st,Boucaud:2002nc,Dudal:2002pq,
Dudal:2003vv,Browne:2003uv,Sobreiro:2004us,Browne:2004mk,
Gracey:2004bk,Dudal:2005na,Li:2004te,Boucaud:2005rm,RuizArriola:2004en,Suzuki:2004dw,
Gubarev:2005it,Furui:2005bu,Chernodub:2005gz,Kekez:2003ri,Kekez:2005ie,Furui:12ks,Boucaud:2005xn,RuizArriola:2006gq,Megias:2005ve}
over the past few years. An OPE argument based on lattice
simulations has provided evidence that this condensate could account
for quadratic power corrections of the form $\sim\frac{1}{Q^2}$,
reported in the running of the coupling constant as well as in the
gluon propagator
\cite{Kondo:2001nq,Boucaud:2001st,Boucaud:2002nc,RuizArriola:2004en,RuizArriola:2006gq,Furui:2005bu,Furui:12ks}.

This nonvanishing condensate $\left\langle A_\mu^2\right\rangle$
gives rise to a dynamically generated gluon mass,
\cite{Verschelde:2001ia,Dudal:2003vv,Kondo:2001nq,Browne:2004mk,Gracey:2004bk}.
The appearance of mass parameters in the gluon two point function is
a common feature of the expressions employed to fit the numerical
data obtained from lattice simulations,
\cite{Marenzoni:1994ap,Leinweber:1998uu,Langfeld:2001cz}. Let us
mention that a gluon mass has been found to be useful also in the
phenomenological context,
\cite{Parisi:1980jy,Field:2001iu}.

The local operator $A_\mu^2$ in the Landau gauge has witnessed a
renewed interest due to the recent works
\cite{Gubarev:2000nz,Gubarev:2000eu}, as the quantity
\begin{equation}\label{intro1}
    \left\langle A^2_{\min}\right\rangle\equiv\min_{U\in SU(N)}\frac{1}{VT}\int
    d^4x \left\langle\left(A_\mu^U\right)^2\right\rangle\;,
\end{equation}
which is gauge invariant due to the minimization along the gauge
orbits, could be physically relevant. In fact, as shown in
\cite{Gubarev:2000nz,Gubarev:2000eu} in the case of compact
three-dimensional QED, the quantity $\left\langle
A^2_{\min}\right\rangle$ seems to be useful in order to detect the
presence of nontrivial field configurations like monopoles. One
should notice that the operator $A_{\min}^2$ is highly nonlocal and
therefore it falls beyond the standard OPE realm that refers to
local operators. One can show that $A_{\min}^2$ can be written as a
infinite series of nonlocal terms, see
\cite{Capri:2005dy,Lavelle:1995ty} and references therein, namely
\begin{eqnarray}
A_{\min }^{2} &=&\frac{1}{2}\int d^{4}x\left[ A_{\mu }^{a}\left( \delta _{\mu \nu }-\frac{%
\partial _{\mu }\partial _{\nu }}{\partial ^{2}}\right) A_{\nu
}^{a}-gf^{abc}\left( \frac{\partial _{\nu }}{\partial ^{2}}\partial
A^{a}\right) \left( \frac{1}{\partial ^{2}}\partial {A}^{b}\right)
A_{\nu }^{c}\right] \;+O(A^{4})\;.  \label{intro2}
\end{eqnarray}
However, in the Landau gauge, $\p_\mu A_\mu=0$, all nonlocal terms
of expression (\ref{intro2}) drop out, so that $A_{\min }^{2}$
reduces to the local operator $A_\mu^2$, hence the interest in the
Landau gauge and its dimension two gluon condensate. However, a
complication, as already outlined in our previous paper
\cite{Capri:2005dy}, is that the explicit determination of the
\emph{absolute} minimum of $A_\mu^2$ along its gauge orbit, and
moreover of its vacuum expectation value, is a very delicate issue
intimately related to the problem of the Gribov copies
\cite{Gribov:1977wm}. We refer to \cite{Capri:2005dy} for some more
explanation and the original references concerning this point.

Nevertheless, some nontrivial results were proven concerning the
operator $A_\mu^2$. In particular, we mention its multiplicative
renormalizability to all orders of perturbation theory, in addition
to an interesting and numerically verified relation concerning its
anomalous dimension \cite{Dudal:2002pq,Gracey:2002yt}. An effective
potential approach consistent with the renormalization group
requirements has also been worked out for this operator, giving
further evidence of a nonvanishing condensate $\left\langle
A_\mu^2\right\rangle\neq0$, which  lowers the nonperturbative vacuum
energy \cite{Verschelde:2001ia}.

A somewhat weak point about the operator $A^2_{\min}$ is that it is
unclear how to deal with it in gauges other than the Landau gauge.
Till now, it seems hopeless to prove its renormalizability out of
the Landau gauge. In fact, at the classical level, adding
(\ref{intro1}) to the Yang-Mills action is equivalent to add the
so-called Stueckelberg action, which is known to be not
renormalizable \cite{Ruegg:2003ps,vanDam:1970vg}. We refer, once
more, to \cite{Capri:2005dy} for details and references.

In recent years, some progress has also been made in the potential
relevance of dimension two condensates beyond the Landau gauge. We
were able to prove the renormalizability of certain local operators
like: $A_\mu^2$ in the linear covariant gauges, $(\frac{1}{2}A_\mu^a
A_\mu^a +\alpha \occ^a c^a)$ in the nonlinear Curci-Ferrari gauges,
and $(\frac{1}{2}A_\mu^\beta A_\mu^\beta +\alpha \occ^\beta
c^\beta)$ in the maximal Abelian gauges \footnote{The color index
$\beta$ runs over the $N(N-1)$ off-diagonal generators of $SU(N)$
and $\alpha$ is a gauge parameter.}. A renormalizable effective
potential for these operators has been constructed, giving rise to a
nontrivial value for the corresponding condensates, and a dynamical
gluon mass parameter emerged in each of these gauges
\cite{Dudal:2003by,Dudal:2003gu,Dudal:2003pe,Dudal:2003np,Dudal:2004rx}.
There also have been attempts to include possible effects of Gribov
copies
\cite{Sobreiro:2004us,Dudal:2005na,Sobreiro:2005vn,Capri:2005tj}.
Unfortunately, the amount of numerical data available from lattice
simulations is rather scarce in the aforementioned gauges.
Nevertheless, let us mention that a dynamical gluon mass in the
maximal Abelian gauge has been reported in
\cite{Amemiya:1998jz,Bornyakov:2003ee}. In the Coulomb gauge too,
the relevance of a dimension two condensate has been touched upon in
the past \cite{Greensite:1985vq}.

Although the renormalizability of the foregoing dimension two
operators is a nontrivial and important fact in its own right, their
lack of gauge invariance is a less welcome feature. Moreover, at
present, it is yet an open question whether these operators might be
related in some way to a gauge invariant gluon mass.

Many aspects of the dimension two condensates and of the related
issue of dynamical mass generation in Yang-Mills theories need
further understanding. An important step forwards would be a gauge
invariant mechanism behind a dynamical mass, without giving up the
important renormalization aspects of quantum field theory.

We set a first step in this direction in our previous paper
\cite{Capri:2005dy}. As a local gauge invariant operator of
dimension two does not exist, and since locality of the action is
almost indispensable to prove renormalizability to all orders and to
have a consistent calculational framework at hand, we could look for
a nonlocal operator that is localizable by introducing an additional
set of fields. As pointed out in \cite{Capri:2005dy},
this task looks extremely difficult for the operator $A^2_{\min}$ if we reckon
that an \emph{infinite} series of nonlocal terms is required, as
displayed in (\ref{intro2}). Instead, we turned our attention to the
gauge invariant operator
\begin{equation}
F\frac{1}{D^2}F\equiv\frac{1}{VT}\int d^{4}xF_{\mu \nu }^{a}\left[
\left( D^{2}\right) ^{-1}\right] ^{ab}F_{\mu \nu }^{b}\;,
\label{intro3}
\end{equation}
where $D^2=D_{\mu}D_{\mu}$ is the covariant Laplacian, $D_{\mu}$
being the adjoint covariant derivative. The operator (\ref{intro3})
already appeared in relation to gluon mass generation in
three-dimensional Yang-Mills theories \cite{Jackiw:1997jg}.

The usefulness of the operator (\ref{intro3}) relies on the fact
that, when it is added to the usual Yang-Mills Lagrangian by means
of $-\frac{1}{4}m^2F\frac{1}{D^2}F$, the resulting action can be
easily cast into a local form by introducing a {\em finite} set of
auxiliary fields \cite{Capri:2005dy}. Starting from that particular
localized action, we succeeded in constructing a gauge invariant
classical action $S_{cl}$ containing the mass parameter $m$,
enjoying renormalizability. This action $S_{cl}$ was identified to
be
\begin{eqnarray}
  S_{phys} &=& S_{cl} +S_{gf}\;,\label{completeaction}\\
  S_{cl}&=&\int d^4x\left[\frac{1}{4}F_{\mu \nu }^{a}F_{\mu \nu }^{a}+\frac{im}{4}(B-\overline{B})_{\mu\nu}^aF_{\mu\nu}^a
  +\frac{1}{4}\left( \overline{B}_{\mu \nu
}^{a}D_{\sigma }^{ab}D_{\sigma }^{bc}B_{\mu \nu
}^{c}-\overline{G}_{\mu \nu }^{a}D_{\sigma }^{ab}D_{\sigma
}^{bc}G_{\mu \nu
}^{c}\right)\right.\nonumber\\
&-&\left.\frac{3}{8}%
m^{2}\lambda _{1}\left( \overline{B}_{\mu \nu }^{a}B_{\mu \nu
}^{a}-\overline{G}_{\mu \nu }^{a}G_{\mu \nu }^{a}\right)
+m^{2}\frac{\lambda _{3}}{32}\left( \overline{B}_{\mu \nu
}^{a}-B_{\mu \nu }^{a}\right) ^{2}+
\frac{\lambda^{abcd}}{16}\left( \overline{B}_{\mu\nu}^{a}B_{\mu\nu}^{b}-\overline{G}_{\mu\nu}^{a}G_{\mu\nu}^{b}%
\right)\left( \overline{B}_{\rho\sigma}^{c}B_{\rho\sigma}^{d}-\overline{G}_{\rho\sigma}^{c}G_{\rho\sigma}^{d}%
\right) \right]\;, \label{completeactionb}\\
S_{gf}&=&\int d^{4}x\;\left( \frac{\alpha }{2}b^{a}b^{a}+b^{a}%
\partial _{\mu }A_{\mu }^{a}+\overline{c}^{a}\partial _{\mu }D_{\mu
}^{ab}c^{b}\right)\;.\label{lcg}
\end{eqnarray}
We notice that we had to introduce a new quartic tensor coupling
$\lambda^{abcd}$, as well as two new mass couplings $\lambda_1$ and
$\lambda_3$. The renormalizability was proven to all orders in the
class of linear covariant gauges, implemented through $S_{gf}$, via
the algebraic renormalization formalism \cite{Piguet:1995er}.
Without the new couplings, i.e. when $\lambda_1\equiv0$,
$\lambda_3\equiv0$, $\lambda^{abcd}\equiv0$, the previous action
would not be renormalizable.

In this paper, we present further results concerning the action
(\ref{completeactionb}) obtained in \cite{Capri:2005dy}. In section
II, we provide a short summary of the construction of the model
(\ref{completeactionb}) and we present a detailed discussion of the
renormalization of the tensor coupling $\lambda^{abcd}$, not given
in \cite{Capri:2005dy}. We draw attention to the existence of an
extended version of the usual nilpotent BRST symmetry for the model
(\ref{completeaction}). We introduce a kind of supersymmetry between
the novel fields
$\left\{B_{\mu\nu}^a,\overline{B}_{\mu\nu}^a,G_{\mu\nu}^a,\overline{G}_{\mu\nu}^a\right\}$
which is enjoyed by the massless version,  $m$~$=$~$0$, of the action
(\ref{completeaction}). In section III, we discuss the explicit
renormalization of several quantities. The fields and the mass $m$
are renormalized to two loop order in the $\MSbar$ scheme. The
$\beta^{abcd}$-function of the tensor coupling $\lambda^{abcd}$ is
determined at one loop, by means of which it shall also become clear
that radiative corrections (re)introduce anyhow the quartic
interaction in the novel fields in the action
(\ref{completeaction}). These fields are thus more than simple
auxiliary fields, which appear at most quadratically. A few internal
checks on the results are included, such as the explicit gauge
parameter independence of the anomalous dimension of $m$. It is also
found that the original Yang-Mills quantities renormalize
identically as when the usual Yang-Mills action would have been
used. This is indicative of the fact that the massless version of
(\ref{completeaction}) might be equivalent to Yang-Mills theory,
quantized in the same gauge. This is a nontrivial statement, due to
the presence of the term proportional to the tensor coupling
$\lambda^{abcd}$. In section IV, we use the aforementioned
supersymmetry to actually prove the equivalence between the massless
version of (\ref{completeaction}) and Yang-Mills theories. In the
concluding section V, we put forward a few suggestions that might be
useful for future research directions.

\section{Survey of the construction of the model and its
renormalizability.}
In this section we present a short summary of
how we came to the construction of our model (\ref{completeaction}) in \cite{Capri:2005dy}. We
shall also point out a few properties of the corresponding action
not explicitly mentioned in \cite{Capri:2005dy}.

\subsection{The model at the classical level.}
We start from the Yang-Mills action $S_{YM}$ supplemented with a
gauge invariant although nonlocal mass operator
\begin{equation}
S_{YM}+S_{\mathcal{O}}\;,  \label{ymop}
\end{equation}
with the usual Yang-Mills action defined by
\begin{equation}
S_{YM}=\frac{1}{4}\int d^{4}xF_{\mu \nu }^{a}F_{\mu \nu }^{a}\;,
\label{ym}
\end{equation}
and with
\begin{equation}
S_{\mathcal{O}}=-\frac{m^{2}}{4}\int d^{4}xF_{\mu \nu }^{a}\left[
\left( D^{2}\right) ^{-1}\right] ^{ab}F_{\mu \nu }^{b}\;.
\label{massop}
\end{equation}
The field strength is given by
\begin{equation}\label{fs1}
    F_{\mu\nu}^a=\p_\mu A_\nu^a - \p_\nu A_\mu^a - gf^{abc}A_\mu^b
    A_\nu^c\;,
\end{equation}
and the adjoint covariant derivative by
\begin{equation}\label{cd1}
    D_\mu^{ab}=\p_\mu\delta^{ab}-gf^{abc}A_\mu^c\;.
\end{equation}
In order to have a consistent calculational framework at the
perturbative level, we need a local action. To our knowledge, it is
unknown how to treat an action like (\ref{ymop}), such as proving
its renormalizability to all orders of perturbation theory. This is
due to the presence of the nonlocal term (\ref{massop}). As we have
discussed in \cite{Capri:2005dy}, the action (\ref{ymop}) can be
localized by introducing a pair of complex bosonic antisymmetric
tensor fields, $\left( B_{\mu \nu }^{a},\overline{B}_{\mu \nu
}^{a}\right) $, and a pair of complex anticommuting antisymmetric
tensor fields, $\left( \overline{G}_{\mu \nu }^{a},G_{\mu \nu
}^{a}\right) $, belonging to the adjoint representation, according
to which
\begin{eqnarray}
e^{-S_{\mathcal{O}}}&=&\int D\overline{B}DBD\oG DG\exp \left[
-\left( \frac{1}{4}\int d^{4}x\overline{B}_{\mu \nu }^{a}D_{\sigma
}^{ab}D_{\sigma }^{bc}B_{\mu \nu }^{c}-\frac{1}{4}\int {%
d^{4}x}\overline{G}_{\mu \nu }^{a}D_{\sigma }^{ab}D_{\sigma
}^{bc}G_{\mu
\nu }^{c}+\frac{im}{4}\int d^{4}x\left( B-\overline{B%
}\right) _{\mu \nu }^{a}F_{\mu \nu }^{a}\right) \right] \;.
\label{loc2}
\end{eqnarray}
It is worth mentioning the special limit $m\equiv0$, in which case
we have in fact introduced nothing more than a rather complicated
unity written as \footnote{We omitted an irrelevant normalization
factor.}
\begin{eqnarray}
\int D\overline{B}DBD\oG DG\exp \left[ -\left( \frac{1}{4}\int
d^{4}x\overline{B}_{\mu \nu }^{a}D_{\sigma
}^{ab}D_{\sigma }^{bc}B_{\mu \nu }^{c}-\frac{1}{4}\int {%
d^{4}x}\overline{G}_{\mu \nu }^{a}D_{\sigma }^{ab}D_{\sigma
}^{bc}G_{\mu \nu }^{c}\right) \right]\equiv 1 \;. \label{loc2bis}
\end{eqnarray}
Hence, we have obtained a local, classical and gauge invariant
action
\begin{equation}
S=S_{YM}+S_{BG}+S_{m}\;,  \label{action1}
\end{equation}
where
\begin{eqnarray}
S_{BG} &=&\frac{1}{4}\int d^{4}x\left( \overline{B}_{\mu \nu
}^{a}D_{\sigma }^{ab}D_{\sigma }^{bc}B_{\mu \nu
}^{c}-\overline{G}_{\mu \nu }^{a}D_{\sigma
}^{ab}D_{\sigma }^{bc}G_{\mu \nu }^{c}\right) \;,  \nonumber \\
S_{m} &=& \frac{im}{4}\int d^{4}x\left( B-\overline{B}\right)
_{\mu \nu }^{a}F_{\mu \nu }^{a}\;. \label{actions2}
\end{eqnarray}
The gauge transformations are given by
\begin{eqnarray}
\delta A_{\mu }^{a} &=&-D_{\mu }^{ab}\omega ^{b}\;,  \nonumber \\
\delta B_{\mu \nu }^{a} &=&gf^{abc}\omega ^{b}B_{\mu \nu }^{c}\;,
\nonumber
\\
\delta \overline{B}_{\mu \nu }^{a} &=&gf^{abc}\omega
^{b}\overline{B}_{\mu \nu }^{c}\;,
\nonumber \\
\delta G_{\mu \nu }^{a} &=&gf^{abc}\omega ^{b}G_{\mu \nu }^{c}\;,
\nonumber
\\
\delta \overline{G}_{\mu \nu }^{a} &=&gf^{abc}\omega
^{b}\overline{G}_{\mu \nu }^{c}\;, \label{gtm}
\end{eqnarray}
with $\omega^a$ parametrizing an arbitrary infinitesimal gauge
transformation, so that
\begin{equation}
\delta S=\delta \left( S_{YM}+S_{BG}+S_{m}\right) =0\;.
\label{gtminv}
\end{equation}
\subsection{The model at the quantum level.}
Evidently, the construction of the classical action (\ref{action1})
is only a first step. We still need to investigate if this action
can be renormalized when the quantum corrections are included. This
highly nontrivial task was treated at length in \cite{Capri:2005dy}
to which we refer the interested reader for background information.
Nevertheless, we shall take some time to explain the main idea as
well as to present a detailed analysis of the quantum corrections
affecting the quartic tensor coupling $\lambda^{abcd}$.

As the quantization of a locally invariant gauge model requires the
fixing of the gauge freedom, we shall employ the linear covariant
gauge fixing from now on, as it was done in \cite{Capri:2005dy}, and
which is imposed via (\ref{lcg}).

To actually discuss the renormalizability of (\ref{action1}), we
found it useful to embed it into a more general class of models
described by the action
\begin{eqnarray}
\Sigma  &=&S_{YM}+S_{gf} +\int {d^{4}x}\left(
\overline{B}_{i}^{a}D_{\mu }^{ab}D_{\mu
}^{bc}B_{i}^{c}-\overline{G}_{i}^{a}D_{\mu }^{ab}D_{\mu
}^{bc}G_{i}^{c}\right)
\nonumber \\
&+&\int d^{4}x\left( \left( \overline{U}_{i\mu \nu }G_{i}^{a}+V_{i\mu \nu }\overline{B}%
_{i}^{a}-\overline{V}_{i\mu \nu }B_{i}^{a}+U_{i\mu \nu
}\overline{G}_{i}^{a}\right) F_{\mu \nu }^{a}+\chi _{1}\left(
\overline{V}_{i\mu \nu }\partial ^{2}V_{i\mu \nu
}-\overline{U}_{i\mu \nu }\partial ^{2}U_{i\mu \nu }\right) \right)   \nonumber \\
&+&\int d^{4}x\chi _{2}\left( \overline{V}_{i\mu \nu }\partial
_{\mu }\partial _{\alpha }V_{i\nu \alpha }-\overline{U}_{i\mu \nu
}\partial _{\mu }\partial _{\alpha }U_{i\nu \alpha }\right) -\int
d^{4}x\zeta \left( \overline{U}_{i\mu \nu }U_{i\mu \nu
}\overline{U}_{j\alpha \beta }U_{j\alpha \beta
}\right.\nonumber\\&+&\left.\overline{V}_{i\mu \nu }V_{i\mu \nu
}\overline{V}_{j\alpha \beta }V_{j\alpha \beta
}-2\overline{U}_{i\mu \nu }U_{i\mu \nu }\overline{V}_{j\alpha
\beta }V_{j\alpha \beta }\right) \;, \label{ass1}
\end{eqnarray}
where use has been made of a composite Lorentz index
$i\equiv(\mu,\nu)$, $i=1\ldots6$, corresponding to a global $U(6)$
symmetry \cite{Capri:2005dy} of the action (\ref{ass1}). The
quantities $V_{i\mu\nu}$, $\overline{V}_{i\mu\nu}$, $U_{i\mu\nu}$
and $\overline{U}_{i\mu\nu}$ are local sources. The identification
between objects carrying indices $i$ and $(\mu,\nu)$ is determined
through
\begin{eqnarray}
\left( B_{i}^{a},\overline{B}_{i}^{a},G_{i}^{a},\overline{G}_{i}^{a}\right) =\frac{1}{2%
}\left( B_{\mu \nu }^{a},\overline{B}_{\mu \nu }^{a},G_{\mu \nu }^{a},\overline{G}%
_{\mu \nu }^{a}\right) \;,  \label{if}\nonumber\\ \left( V_{i\mu \nu
},\overline{V}_{i\mu \nu },U_{i\mu \nu },\overline{U}_{i\mu \nu
}\right) =\frac{1}{2}\left( V_{\sigma \rho \mu \nu
},\overline{V}_{\sigma \rho \mu \nu },U_{\sigma \rho \mu \nu
},\overline{U}_{\sigma \rho \mu \nu }\right) \;.
 \label{1q}
\end{eqnarray}
The free parameters $\chi_1$, $\chi_2$ and $\zeta$ are needed for
renormalizability purposes. As far as we are considering Green
functions of elementary fields, their role is irrelevant as they
multiply terms which are polynomial in the external sources.

The reason for introducing the action (\ref{ass1}) is that for
$m=0$, the action (\ref{action1}) enjoys a few global symmetries
which are lost for $m\neq0$, whereas the action (\ref{ass1}) also
enjoys these symmetries when the global symmetry transformations are
suitably extended to the sources. We refer to \cite{Capri:2005dy}
for the details. Evidently, the general action (\ref{ass1}) must
possess the action (\ref{action1}) we are interested in as a special
case. The reader can check that the connection is made by
considering the ``physical'' limit
\begin{eqnarray}
\lim_{phys}\overline{V}_{\sigma \rho \mu \nu }
&=&\lim_{phys}V_{\sigma \rho \mu \nu }=\frac{-im}{2}\left( \delta
_{\sigma \mu }\delta _{\rho \nu }-\delta _{\sigma \nu }\delta _{\rho
\mu }\right) \;,
\nonumber \\
\lim_{phys}\overline{U}_{\sigma \rho \mu \nu
}&=&\lim_{phys}U_{\sigma \rho \mu \nu }=0\;, \label{plm}
\end{eqnarray}
i.e.
\begin{eqnarray}
\lim_{phys}\Sigma =S \label{plm2}\,.
\end{eqnarray}
In \cite{Capri:2005dy}, it was shown that the action (\ref{ass1})
obeys a large set of Ward identities. We shall not list them here,
but mention that the action $\Sigma$ is invariant under a nilpotent
BRST transformation $s$, acting on the fields as
\begin{eqnarray}\label{brst1}
sA_{\mu }^{a} &=&-D_{\mu }^{ab}c^{b}\;,  \nonumber \\
sc^{a} &=&\frac{g}{2}f^{abc}c^{b}c^{c}\;,  \nonumber \\
sB_{\mu \nu }^{a} &=&gf^{abc}c^{b}B_{\mu \nu }^{c}+G_{\mu \nu
}^{a}\;,
\nonumber \\
s\overline{B}_{\mu \nu }^{a} &=&gf^{abc}c^{b}\overline{B}_{\mu \nu
}^{c}\;,
\nonumber\\
sG_{\mu \nu }^{a} &=&gf^{abc}c^{b}G_{\mu \nu }^{c}\;,  \nonumber \\
s\overline{G}_{\mu \nu }^{a} &=&gf^{abc}c^{b}\overline{G}_{\mu \nu
}^{c}+\overline{B}_{\mu
\nu }^{a}\;,  \nonumber \\
s\overline{c}^{a} &=&b^{a}\;,  \nonumber \\
sb^{a} &=&0\;, \end{eqnarray} and on the sources as
\begin{eqnarray}\label{brst2}
sV_{i\mu\nu}&=&U_{i\mu\nu}\;,\nonumber\\
sU_{i\mu\nu}&=&0\;,\nonumber\\
s\overline{U}_{i\mu\nu}&=&\overline{V}_{i\mu\nu}\;,\nonumber\\
s\overline{V}_{i\mu\nu}&=&0\;,
\end{eqnarray}
such that
\begin{eqnarray}
s\Sigma&=&0\;,\nonumber\\
 s^{2} &=&0\;.  \label{bi}
\end{eqnarray}
By employing the algebraic renormalization technique
\cite{Piguet:1995er}, it was found that (\ref{ass1}) is not yet the
most general action compatible with all the Ward identities,
including the Slavnov-Taylor identity associated to the BRST
invariance described by (\ref{bi}). The most general and
renormalizable action was identified to be
\begin{equation}\label{renact}
S_{gen}=\Sigma+S_{\lambda}\;,
\end{equation}
where
\begin{eqnarray}
S_{\lambda } &=&\int d^{4}x\left[ \lambda _{1}\left( \overline{B}%
_{i}^{a}B_{i}^{a}-\overline{G}_{i}^{a}G_{i}^{a}\right) \left(
\overline{V}_{j\mu \nu }V_{j\mu \nu }-\overline{U}_{j\mu \nu
}U_{j\mu \nu
}\right) +\frac{\lambda^{abcd}}{16}\left( \overline{B}_{i}^{a}B_{i}^{b}-\overline{G}_{i}^{a}G_{i}^{b}%
\right)\left( \overline{B}_{j}^{c}B_{j}^{d}-\overline{G}_{j}^{c}G_{j}^{d}%
\right) \right.
\nonumber \\
&+&\left. \lambda _{3}\left( \overline{B}_{i}^{a}G_{j}^{a}V_{i\mu \nu }\overline{U}%
_{j\mu \nu }+\overline{G}_{i}^{a}G_{j}^{a}U_{i\mu \nu }\overline{U}_{j\mu \nu }+\overline{B}%
_{i}^{a}B_{j}^{a}V_{i\mu \nu }\overline{V}_{j\mu \nu }-\overline{G}%
_{i}^{a}B_{j}^{a}U_{i\mu \nu }\overline{V}_{j\mu \nu }\right.
\right. \nonumber
\\
&-&\left. \left. G_{i}^{a}B_{j}^{a}\overline{U}_{i\mu \nu }\overline{V}_{j\mu \nu }+%
\overline{G}_{i}^{a}\overline{B}_{j}^{a}U_{i\mu \nu }V_{j\mu \nu }-\frac{1}{2}%
B_{i}^{a}B_{j}^{a}\overline{V}_{i\mu \nu }\overline{V}_{j\mu \nu }+\frac{1}{2}%
G_{i}^{a}G_{j}^{a}\overline{U}_{i\mu \nu }\overline{U}_{j\mu \nu
}\right. \right.
\nonumber \\
&-&\left. \left.
\frac{1}{2}\overline{B}_{i}^{a}\overline{B}_{j}^{a}V_{i\mu \nu
}V_{j\mu \nu
}+\frac{1}{2}\overline{G}_{i}^{a}\overline{G}_{j}^{a}U_{i\mu \nu
}U_{j\mu \nu }\right) \right] \;. \label{l1}
\end{eqnarray}
The quantities $\lambda_1$ and $\lambda_3$ are independent scalar
couplings, whereas $\lambda^{abcd}$ is an invariant rank 4 tensor
coupling, obeying the generalized Jacobi identity
\begin{equation}\label{jacobigen}
    f^{man}\lambda^{mbcd}+f^{mbn}\lambda^{amcd}+f^{mcn}\lambda^{abmd}+f^{mdn}\lambda^{abcm}=0\,,
\end{equation}
and subject to the following symmetry constraints
\begin{eqnarray}
\lambda^{abcd}=\lambda^{cdab} \;, \nonumber \\
\lambda^{abcd}=\lambda^{bacd} \;, \label{abcd}
\end{eqnarray}
which can be read off from the vertex that $\lambda^{abcd}$
multiplies. When we specify the action (\ref{renact}) to the
physical values (\ref{plm}), we obtain the main outcome of the
paper \cite{Capri:2005dy}, which is the physical action $S_{phys}$
given in (\ref{completeaction}), that is renormalizable to all
orders of perturbation theory in the linear covariant gauge,
imposed via $S_{gf}$. The renormalizability is of course ensured
as (\ref{completeaction}) is a special case of the more general
renormalizable action (\ref{renact}), since
\begin{equation}\label{relatie}
    \lim_{phys}S_{gen}=S_{phys}\;.
\end{equation}
We notice that the couplings $\lambda_1$ and $\lambda_3$ are in fact
part of the mass matrix of the fields
$\left\{B_{\mu\nu}^a,\oB_{\mu\nu}^a,G_{\mu\nu}^a,\oG_{\mu\nu}^a\right\}$.

We end this subsection by mentioning that the classical action
$S_{cl}$ is also invariant with respect to the gauge transformations
(\ref{gtm}), since the terms
$\propto\left\{\lambda_{1},\lambda_{3},\lambda^{abcd}\right\}$ are
separately gauge invariant.

\subsection{The renormalization of the tensor coupling $\lambda^{abcd}$.}
As already mentioned, this section is devoted to providing further
details of the renormalization of the tensor coupling
$\lambda^{abcd}$, not fully covered in \cite{Capri:2005dy}.

The term we are interested in at the level of the bare
\footnote{Bare quantities are denoted with a subscript ``$o$''.}
action is given by
\begin{equation}\label{b1}
    \int d^4x\left[\frac{\lambda_o^{abcd}}{16}\left( \overline{B}_{i,o}^{a}B_{i,o}^{b}-\overline{G}_{i,o}^{a}G_{i,o}^{b}%
\right)\left(
\overline{B}_{j,o}^{c}B_{j,o}^{d}-\overline{G}_{j,o}^{c}G_{j,o}^{d}
\right)\right]\;,
\end{equation}
where, in the notation of \cite{Capri:2005dy}
\begin{equation}\label{b2}
\left\{B,\overline{B},G,\overline{G}\right\}_{o,i}^{a}=\sqrt{Z_b}\left\{B,\overline{B},G,\overline{G}\right\}_{i}^{a}
=\left[1+\eta\left(a_3+\frac{1}{2}a_4\right)\right]\left\{B,\overline{B},G,\overline{G}\right\}_{i}^{a}\;,
\end{equation}
where $a_3, a_4$ are arbitrary coefficients and $\eta$ stands for a
perturbative expansion parameter \cite{Capri:2005dy}.

The most general counterterm corresponding to
the renormalization of the 4-point vertex $\left( \overline{B}_{i}^{a}B_{i}^{b}-\overline{G}_{i}^{a}G_{i}^{b}%
\right)\left( \overline{B}_{j}^{c}B_{j}^{d}-\overline{G}_{j}^{c}G_{j}^{d}%
\right)$ turns out to be given by
\begin{equation}\label{b4a}
    (4a_3+\tilde{a}_6)\frac{\mathcal{M}^{abcd}}{16}\left( \overline{B}_{i}^{a}B_{i}^{b}-\overline{G}_{i}^{a}G_{i}^{b}\right)\left( \overline{B}_{j}^{c}B_{j}^{d}-\overline{G}_{j}^{c}G_{j}^{d}%
\right)\;,
\end{equation}
where $\tilde{a}_6$ is a free coefficient and $\mathcal{M}^{abcd}$
is an arbitrary invariant rank 4 tensor, composed of all the other
available tensors (such as $\lambda^{abcd}$, $\delta^{ab}$,
$f^{abc}$ and invariant objects constructed from these and $T^a$).
By the Ward identities, it is nevertheless restricted by
\begin{eqnarray}\label{b4}
   \mathcal{M}^{abcd}&=& \mathcal{M}^{cdab}\,, \nonumber\\
  \mathcal{M}^{abcd} &=& \mathcal{M}^{bacd}\;,
\end{eqnarray}
which are of course the same symmetry constraints as those for
$\lambda^{abcd}$, see (\ref{abcd}). Also the Jacobi identity
(\ref{jacobigen}) applies to $\mathcal{M}^{abcd}$.

The counterterm is thus not necessarily directly proportional to the
original tensor $\lambda^{abcd}$. This has a simple diagrammatical
explanation, as diagrams contributing to
the 4-point interaction $\propto \left( \overline{B}_{i}^{a}B_{i}^{b}-\overline{G}_{i}^{a}G_{i}^{b}%
\right)\left( \overline{B}_{j}^{c}B_{j}^{d}-\overline{G}_{j}^{c}G_{j}^{d}%
\right)$ can be constructed with the other available interactions.
This also means that, even if $\lambda^{abcd}=0$, radiative
corrections shall reintroduce this 4-point interaction. This shall
become more clear in section III, where the explicit results are
discussed.

We can thus decompose the bare tensor coupling $\lambda_o^{abcd}$ as
\begin{equation}\label{b5}
    \lambda_o^{abcd}=Z\lambda^{abcd}+Z^{abcd}\;,
\end{equation}
where $Z$ and $Z^{abcd}$ contain the counterterm information, more
precisely $Z$ contains the counterterm information directly
proportional to $\lambda^{abcd}$, while $Z^{abcd}$ contains, so to
say, all the rest. Evidently, $Z^{abcd}$ will obey analogous
constraints as given in (\ref{abcd}) or (\ref{b4}). The tensor
$\mathcal{M}^{abcd}$ can be decomposed similarly into
\begin{equation}\label{b7}
    \mathcal{M}^{abcd}=\lambda^{abcd}+\underbrace{\mathcal{M}^{abcd}-\lambda^{abcd}}_{\equiv
    \mathcal{N}^{abcd}}\;.
\end{equation}
In the previous paper \cite{Capri:2005dy}, we erroneously omitted
the $\mathcal{N}^{abcd}$ part. Using (\ref{b4a}), (\ref{b5}) and
(\ref{b7}) allows for a simple identification, being
\begin{eqnarray}\label{234}
    Z &=& 1+\eta(\tilde{a}_6-2a_4)\;,\nonumber\\
    Z^{abcd}&=&\eta(4a_3+\widetilde{a}_6)\mathcal{N}^{abcd}\;.
\end{eqnarray}
Consequently, the model is still renormalizable to all orders,
although $\lambda^{abcd}$ is not multiplicatively renormalizable in
the naive sense. The situation can be directly compared with Higgs
inspired models like the Coleman-Weinberg action
\cite{Coleman:1973jx,Knecht:2001cc}, in the sense that also there a
similar mixing occurs between the different couplings, in casu the
gauge coupling $e^2$ and the Higgs coupling $\lambda$. This is
nicely reflected in the $\beta$-functions for the couplings, which
are series in both $e^2$ and $\lambda$. It is even so that setting
the Higgs coupling $\lambda\equiv0$ does not make
$\beta_\lambda(e^2,\lambda)$ vanish. See \cite{Knecht:2001cc} for
the three loop expressions. As we shall discuss later in this paper,
the $\beta^{abcd}$-function of the tensor coupling $\lambda^{abcd}$
will be influenced by the gauge coupling $g^2$. Vice versa, one
might expect that $\lambda^{abcd}$ could enter, in a suitable
colorless combination, the $\beta_{g^2}$-function for $g^2$. This is
however not the case. We shall present the general argument behind
this in section IV. The $\beta_{g^2}$-function remains thus
identical to the well known $\beta(g^2)$-function of Yang-Mills
theory.

Let us end this subsection by mentioning that the method of using
the extended action (\ref{renact}), which is a generalization of
another action like (\ref{completeaction}) and which exhibits a
larger set of Ward identities, turns out to be a powerful tool in
order to establish renormalizability to all orders. This is
reminiscent of Zwanziger's approach to prove the renormalizability
of a local action describing the restriction to the first Gribov
horizon \cite{Zwanziger:1989mf,Zwanziger:1992qr}.

\subsection{A few words about the BRST symmetry and a kind of supersymmetry.}
Let us now return for a moment to the action $S_{phys}$ given in
(\ref{completeaction}). As it is a gauge fixed action, we expect
that it should have a nilpotent BRST symmetry. However, one shall
easily recognize that the BRST transformation $s$ as defined in
(\ref{brst1}) no longer constitutes a symmetry of the action
(\ref{completeaction}). This is due to the fact that setting the
sources to their physical values (\ref{plm}) breaks the BRST $s$ as
the transformations (\ref{brst2}) are incompatible with the desired
physical values (\ref{plm}).

Let us take a closer look at the breaking of this BRST
transformation $s$. Let us define another transformation
$\widetilde{s}$ at the level of the fields by
\begin{eqnarray}
\widetilde{s} A_{\mu }^{a} &=&-D_{\mu }^{ab}c ^{b}\;,  \nonumber \\
\widetilde{s} c^{a} &=&\frac{g}{2}f^{abc}c^ac ^{b}\;,  \nonumber \\
 \widetilde{s} B_{\mu \nu }^{a} &=&gf^{abc}c ^{b}B_{\mu \nu }^{c}\;,
\nonumber
\\
\widetilde{s} \overline{B}_{\mu \nu }^{a} &=&gf^{abc}c
^{b}\overline{B}_{\mu \nu }^{c}\;,
\nonumber \\
\widetilde{s} G_{\mu \nu }^{a} &=&gf^{abc}c ^{b}G_{\mu \nu }^{c}\;,
\nonumber
\\
\widetilde{s} \overline{G}_{\mu \nu }^{a} &=&gf^{abc}c
^{b}\overline{G}_{\mu
\nu }^{c}\;,\nonumber\\
\widetilde{s}\overline{c}^{a} &=&b^a\;,  \nonumber \\
\widetilde{s} b^{a} &=&0\;.
 \label{brst3}
\end{eqnarray}
A little algebra yields
\begin{eqnarray}\label{brst4}
% \nonumber to remove numbering (before each equation)
  \widetilde{s}S_{phys} &=& 0\;, \\
  \widetilde{s}^2 &=& 0\,.
\end{eqnarray}
Hence, the action $S_{phys}$ is invariant with respect to a
nilpotent BRST transformation $\widetilde{s}$. We obtained thus a
gauge field theory, described by the action $S_{phys}$,
(\ref{completeaction}), containing a mass term, and which has the
property of being renormalizable, while nevertheless a nilpotent
BRST transformation expressing the gauge invariance after gauge
fixing exists simultaneously. It is clear that $\widetilde{s}$
stands for the usual BRST transformation, well known from
literature, on the original Yang-Mills fields, whereas the gauge
fixing part $S_{gf}$ given in (\ref{lcg}) can be written as a
$\widetilde{s}$-variation, ensuring that the gauge invariant
physical operators shall not depend on the choice of the gauge
parameter \cite{Piguet:1995er}.

We can relate $\widetilde{s}$ and $s$. Let us start from the
original localized action (\ref{action1}) and let us set $m\equiv0$.
Then it enjoys a nilpotent ``supersymmetry'' between the auxiliary
tensor fields $\left\{B_{\mu\nu}^a, \oB_{\mu\nu}^a, G_{\mu\nu}^a,
\oG_{\mu\nu}^a\right\}$, more precisely if we define the
(anticommuting) transformation $\delta_s$ as
\begin{eqnarray}\label{ss}
\delta_s B_{\mu\nu}^a &=& G_{\mu\nu}^a \;,\qquad \delta_s G_{\mu\nu}^a =0 \;,\nonumber\\
\delta_s \overline{G}_{\mu\nu}^a &=& \overline{B}_{\mu\nu}^a \;,
\qquad \delta_s \overline{B}_{\mu\nu}^a = 0 \;,\nonumber\\[3mm]
\delta_s \Psi&=&0 \textrm{ for all other fields }\Psi\;,
\end{eqnarray}
then one can check that
\begin{eqnarray}\label{ss2}
    \delta_s^2&=&0\;,\nonumber\\
    \delta_s \left(\left.S\right|_{m=0}\right)&=&0\;.
\end{eqnarray}
Let us mention for further use that, $\delta_s$ being a nilpotent
operator, it possesses its own cohomology, which is easily
identified with polynomials in the original Yang-Mills fields
$\{A^a_{\mu}, b^a, c^a, {\overline c}^a \}$. The auxiliary tensor
fields, $\{B_{\mu\nu}^a,{\overline B}_{\mu\nu}^a, G_{\mu\nu}^a,
{\overline G}_{\mu\nu}^a\}$, do not belong to the cohomology of
$\delta_s$, as a consequence of the fact that they form pairs of
doublets \cite{Piguet:1995er}.

Taking a closer look upon eqns. (\ref{brst1}), (\ref{brst3}) and
(\ref{ss}), one immediately verifies that
\begin{equation}\label{ss3}
    s=\widetilde{s}+\delta_s\;.
\end{equation}
When $m\neq0$, the action (\ref{action1}) is no longer
$\delta_s$-invariant. Nevertheless, this $\delta_s$-symmetry can be
kept if the more general action (\ref{ass1}) is employed, when we
extend the $\delta_s$-invariance to the introduced sources
\footnote{In fact $\delta_s\equiv s$,$\widetilde{s}\equiv0$ for the
set of sources $\left\{U_{i\mu\nu},\overline{U}_{i\mu\nu},
V_{i\mu\nu},\overline{V}_{i\mu\nu}\right\}$. These sources are gauge
(BRST $\widetilde{s}$) singlets. }  as
\begin{eqnarray}\label{ss4}
\delta_sV_{i\mu\nu}&=&U_{i\mu\nu}\;,\nonumber\\
\delta_sU_{i\mu\nu}&=&0\;,\nonumber\\
\delta_s\overline{U}_{i\mu\nu}&=&\overline{V}_{i\mu\nu}\;,\nonumber\\
\delta_s\overline{V}_{i\mu\nu}&=&0\;.
\end{eqnarray}
Eventually, the most general and renormalizable action
(\ref{renact}) turns out to be compatible with the
$\delta_s$-invariance too, as it should be. This is obvious if we
recognize that we can write
\begin{eqnarray}
S_{gen} &=&S_{YM}+S_{gf}+
\delta_s\int d^4x \oG_{i}^aD_{\sigma}^{ab}D_{\sigma}^{bc}B_i^c\nonumber\\
&+&\delta_s\int d^{4}x\left[ \left( V_{i\mu \nu }\overline{G}_{i}^{a}-\overline{U}%
_{i\mu \nu }B_{i}^{a}\right) F_{\mu \nu }^{a}+\chi
_{1}\overline{U}_{i\mu \nu
}\partial ^{2}V_{i\mu \nu }\right.   \nonumber \\
 &+&\left.\chi _{2}\overline{U}_{i\mu \nu }\partial _{\mu }\partial _{\alpha
}V_{i\nu \alpha }-\zeta \left( \overline{U}_{i\mu \nu }V_{i\mu \nu }\overline{V}%
_{j\alpha \beta }V_{j\alpha \beta }-\overline{U}_{i\mu \nu }V_{i\mu \nu }\overline{U}%
_{j\alpha \beta }U_{j\alpha \beta }\right) \right] \nonumber\\
&+&\delta_s\int d^{4}x\left[\vphantom{\frac{1}{2}} \lambda
_{1}\left(B_i^a\overline{G}_i^a\left(\overline{V}_{j\mu\nu}V_{j\mu\nu}-\overline{U}_{j\mu\nu}U_{j\mu\nu}\right)\right)\right.\nonumber\\
&+&\left.\lambda^{abcd}\left(\left( \overline{B}_{i }^{a}B_{i}^{b}-\overline{G}_{i }^{a}G_{i }^{b}%
\right)\left(\overline{G}_{j}^{c}B_{j}^c\right)\right)
+\lambda_3\left(B_j^a\overline{U}_{j\mu\nu}\left(\overline{B}_i^aV_{i\mu\nu}-\overline{G}_i^aU_{i\mu\nu}
\right)\right)\right.\nonumber\\
&-&\left.\frac{1}{2}\lambda_3\left(B_j^a
\overline{U}_{j\mu\nu}\left(G_i^a\overline{U}_{i\mu\nu}+B_i^a\overline{V}_{i\mu\nu}\right)\right)+\frac{1}{2}\lambda_3\left(\overline{G}_j^a
V_{j\mu\nu}\left(\overline{G}_i^a
U_{i\mu\nu}-\overline{B}_i^aV_{i\mu\nu}\right)\right)\right] \;,
\label{r13}
\end{eqnarray}
and invoke the nilpotency of $\delta_s$.

What happens when we return to the physical action
(\ref{completeaction})? Clearly, the $\delta_s$-invariance is broken
as (\ref{plm}) and (\ref{ss4}) are incompatible. The presence of the
mass $m$ thus breaks the supersymmetry $\delta_s$. As a consequence,
by keeping (\ref{ss3}) in mind for the fields, the BRST
tranformation $s$ is lost too. Fortunately, we recover another BRST
invariance $\widetilde{s}$, (\ref{brst3}), for the physical action.
We shall come back to the relevance and use of the
$\delta_s$-supersymmetry in section IV.

\subsection{Intermediate conclusion.}
The classical gauge invariant action $S_{cl}$ can be quantized in
the linear covariant gauges, whereby a nilpotent BRST symmetry and
renormalizability to all orders of perturbation theory are present.

The most famous gauge models exhibiting renormalizability with the
possibility of massive gauge bosons are of course those based on the
Higgs mechanism, which is related to a spontaneous gauge symmetry
breaking
\cite{Higgs:1964pj,Higgs:1964ia,Higgs:1966ev,Englert:1964et,Guralnik:1964eu}.

Few other Yang-Mills models exhibiting mass terms for the gauge
bosons exist. We mention those based on the Stueckelberg formalism,
which give rise to a nonpolynomial action in the extra Stueckelberg
fields. However, these models lack renormalizability
\cite{Ruegg:2003ps,vanDam:1970vg}. Other models are based on the
works \cite{Curci:1976bt,Curci:1976ar} by Curci and Ferrari.
Although the resulting models are renormalizable, they do not have a
classical gauge invariant counterpart, since the mass terms that are
allowed/needed by renormalizability are not gauge invariant terms.
Typically, the mass term is of the form $\frac{1}{2}m^2\left(A_\mu^a
A_\mu^a +\alpha \occ^a c^a\right)$, where $\alpha$ is the gauge
parameter \footnote{When $m\equiv0$, these models are massless
Yang-Mills theories fixed in the Curci-Ferrari gauge with $\alpha$
the associated gauge parameter.}. Next to the Curci-Ferrari gauges,
the special case of the Landau gauge, corresponding to taking the
limit $\alpha=0$ for the Curci-Ferrari gauge parameter, and the
maximal Abelian gauges can also be used to built up such
renormalizable massive models. Unfortunately, these models have the
problem of being not unitary \cite{Ojima:1981fs,deBoer:1995dh}, a
fact related to the lack of a nilpotent BRST transformation
\footnote{There is a BRST symmetry, but it is not nilpotent.}.
Nevertheless, in the past few years a lot of interest arose in these
dimension two operators from the viewpoint of massless Yang-Mills
theories quantized in a specific gauge. As these operators turn out
to be renormalizable to all orders of perturbation theory in the
specific gauge chosen
\cite{Dudal:2003np,Dudal:2003pe,Dudal:2004rx,Dudal:2002pq}, a
consistent framework can be constructed to investigate the
condensation of these renormalizable albeit non gauge invariant
operators
\cite{Verschelde:2001ia,Dudal:2003by,Dudal:2003gu,Dudal:2003vv,Dudal:2004rx,Browne:2003uv,Dudal:2005na,Sobreiro:2004us}.
This has resulted in a dynamical mass generating mechanism in gauge
fixed Yang-Mills theories
\cite{Dudal:2003by,Dudal:2003gu,Dudal:2003vv,Dudal:2004rx,Browne:2003uv,Dudal:2005na,Kondo:2001nq,Browne:2004mk,Gracey:2004bk}.

\section{Two loop calculations.}
We now detail the actual computation of the two loop anomalous
dimension of the fields and the one loop $\beta$-function of the
tensor coupling $\lambda^{abcd}$. In order to deduce the
renormalization group functions, there are two possible ways to
proceed. One is to regard the extra gluon mass operator as part of
the free Lagrangian and work with completely massive gluon and
localizing ghosts throughout. It transpires that this would be
extremely tedious for various reasons. First, although the
propagators will be massive there will be a $2$-point mixing between
the gluon and localizing ghosts leading to a mixed propagator.
Whilst it is possible to handle such a situation, as has recently
been achieved in a similar localization in \cite{Gracey:2005cx}, it
requires a significantly large number of Feynman diagrams to perform
the full renormalization. Moreover, one needs to develop an
algorithm to systematically integrate massive Feynman diagrams where
the masses are in principle all divergent.  Although algorithms have
been developed for similar but simpler renormalizations, we do not
pursue this avenue here mainly because the extra effort for this
route is not necessary given that there is a simpler alternative.
This is to regard the mass operator as an insertion and split the
Lagrangian into a free piece involving massless fields with the
remainder being transported to the interaction Lagrangian. Hence to
renormalize the operator will involve its insertion into a massless
Green function, after the fields and couplings have been
renormalized in the massless Lagrangian. This is possible since it
has been demonstrated that the ultraviolet structure of the
renormalization constants remain unchanged in $\MSbar$ whether the
gluon mass operator is present or not \cite{Capri:2005dy}. Moreover,
given that the massless field approach is simpler and more
attractive, we can use the {\sc Mincer} algorithm to perform the
actual computations. This algorithm, \cite{Gorishnii:1989gt},
written in the symbolic manipulation language {\sc Form},
\cite{Vermaseren:2000nd,Larin:1991fz}, is devised to extract the
divergences from massless $2$-point functions. Therefore, it is
ideally suited to deduce the anomalous dimensions of the fields.
Hence we note that for the computations the propagators of the
massless fields in an arbitrary linear covariant gauge are,
\cite{Capri:2005dy},
\begin{eqnarray}
\langle A^a_\mu(p) A^b_\nu(-p) \rangle &=& -~
\frac{\delta^{ab}}{p^2} \left[
\delta_{\mu\nu} ~-~ (1-\alpha) \frac{p_\mu p_\nu}{p^2} \right]\;, \nonumber \\
\langle c^a(p) {\bar c}^b(-p) \rangle &=& \frac{\delta^{ab}}{p^2}
~~~,~~~
\langle \psi(p) {\overline{ \psi}}(-p) \rangle ~=~ \frac{\pslash}{p^2}\;, \nonumber \\
\langle B^a_{\mu\nu}(p) {\bar B}^b_{\sigma\rho}(-p) \rangle &=& -~
\frac{\delta^{ab}}{2p^2} \left[ \delta_{\mu\sigma} \delta_{\nu\rho}
~-~
\delta_{\mu\rho} \delta_{\nu\sigma} \right] \;,\nonumber \\
\langle G^a_{\mu\nu}(p) {\bar G}^b_{\sigma\rho}(-p) \rangle &=& -~
\frac{\delta^{ab}}{2p^2} \left[ \delta_{\mu\sigma} \delta_{\nu\rho}
~-~ \delta_{\mu\rho} \delta_{\nu\sigma} \right]\;,
\end{eqnarray}
where $p$ is the momentum. Using {\sc Qgraf},
\cite{Nogueira:1991ex}, to generate the two loop Feynman diagrams we
have first checked that the {\em same} two loop anomalous dimensions
emerge for the gluon, Faddeev-Popov ghost and quarks in an arbitrary
linear covariant gauge as when the extra localizing ghosts are
absent. This is primarily due to the fact that the $2$-point
functions of the fields do not involve any extra ghosts except
within diagrams. Then they appear with equal and opposite signs due
to the anticommutativity of the $G^a_{\mu\nu}$ ghosts and hence
cancel. This observation shall be given an explicit proof in
section IV. We thus note that the expressions obtained for the
renormalization group functions are the same as the two loop
$\MSbar$ results of
\cite{Jones:1974mm,Caswell:1974gg,Tarasov:1976ef,Egorian:1978zx}.
For the localizing ghosts there is the added feature that the
properties of the $\lambda^{abcd}$ couplings have to be used, as
specified in (\ref{jacobigen}) and (\ref{abcd}). We have implemented
these properties in a {\sc Form} module. However, we note that in
the renormalization of both localizing ghosts, we have assumed that
\begin{equation}
\lambda^{acde} \lambda^{bcde} ~=~ \frac{1}{\NA} \delta^{ab}
\lambda^{pqrs} \lambda^{pqrs} ~~~,~~~ \lambda^{acde} \lambda^{bdce}
~=~ \frac{1}{\NA} \delta^{ab} \lambda^{pqrs} \lambda^{prqs}\;,
\end{equation}
which follows from the fact that there is only one rank two
invariant tensor in a {\em classical} Lie group. If this is not
satisfied then one would require a $2$-point counterterm involving
the $\lambda^{abcd}$ couplings which was not evident in the
algebraic renormalization technology which established the
renormalizability of the localized operator. Hence, at two loops in
$\MSbar$ we find that
\begin{eqnarray}
\gamma_B(a,\lambda) ~=~ \gamma_G(a,\lambda) &=& ( \alpha - 3 ) a ~+~
\left[ \left( \frac{\alpha^2}{4} + 2 \alpha - \frac{61}{6} \right)
C_A^2 ~+~ \frac{10}{3} T_F \Nf \right] a^2  +~ \frac{1}{128N_A}
\lambda^{abcd} \lambda^{acbd}\;, \label{gammab}
\end{eqnarray}
where $N_A$ is the dimension of the adjoint representation of the
colour group, $a$~$=$~$g^2/(16\pi^2)$ and we have also absorbed a
factor of $1/(4\pi)$ into $\lambda^{abcd}$ here and in later
anomalous dimensions. These anomalous dimensions are consistent with
the general observation that these fields must have the {\em same}
renormalization constants, in full agreement with the output of the
Ward identities \cite{Capri:2005dy}.

In order to verify that (\ref{gammab}) is in fact correct, we have
renormalized the $3$-point gluon $B^a_{\mu\nu}$ vertex. Since the
coupling constant renormalization is unaffected by the extra
localizing ghosts (and we have checked this explicitly by
renormalizing the gluon quark vertex), then we can check that the
{\em same} gauge parameter independent coupling constant
renormalization constant emerges from gluon $B^a_{\mu\nu}$ vertex.
Computing the $7$ one loop and $166$ two loop Feynman diagrams it is
reassuring to record that the vertex is finite with the already
determined two loop $\MSbar$ field and coupling constant
renormalization constants. Prior to considering the operator itself,
we need to determine the one loop $\beta$-function for the
$\lambda^{abcd}$ couplings. As this is present in a quartic
interaction it means that to deduce its renormalization constant, we
need to consider a $4$-point function. However, in such a situation
the {\sc Mincer} algorithm is not applicable since two external
momenta have to be nullified and this will lead to spurious infrared
infinities which could potentially corrupt the renormalization
constant. Therefore, for this renormalization only, we have resorted
to using a temporary mass regularization introduced into the
computation using the algorithm of \cite{Chetyrkin:1997fm} and
implemented in {\sc Form}. Consequently, we find the gauge parameter
independent renormalization
\begin{eqnarray}\label{lambdaZ}
\lambda_o^{abcd} &=& \lambda^{abcd} ~+~ \left[ \frac{1}{8} \left(
\lambda^{abpq} \lambda^{cpdq} + \lambda^{apbq} \lambda^{cdpq} +
\lambda^{apcq} \lambda^{bpdq} + \lambda^{apdq} \lambda^{bpcq}
\right) \right.
\nonumber \\
&& \left. -~ 6 C_A \lambda^{abcd} a ~+~ 4 C_A f^{abp} f^{cdp} a^2
~+~ 8 C_A f^{adp} f^{bcp} a^2 ~+~ 48 d_A^{abcd} a^2 \right]
\frac{1}{\varepsilon}\;,
\end{eqnarray}
from both the $\lambda^{abcd} \overline{B}^a_{\mu\nu} B^{b\,\mu\nu}
\overline{B}^c_{\sigma\rho} B^{d\,\sigma\rho}$ and $\lambda^{abcd}
\overline{B}^a_{\mu\nu} B^{b\,\mu\nu} \overline{G}^c_{\sigma\rho}
G^{d\,\sigma\rho}$ vertices where $d_A^{abcd}$ is the totally
symmetric rank four tensor defined by
\begin{equation}
d_A^{abcd} ~=~ \mbox{Tr} \left( T^a_A T^{(b}_A T^c_A T^{d)}_A
\right)\;,
\end{equation}
with $T^a_A$ denoting the group generator in the adjoint
representation, \cite{vanRitbergen:1998pn}. Dimensional
regularization in $d=4-2\varepsilon$ dimensions is used throughout
this paper. Producing the same expression for both these $4$-point
functions, aside from the gauge independence, is a strong check on
their correctness as well as the correct implementation of the group
theory. Unlike for the gauge coupling and its $\beta$-function, the
$\lambda^{abcd}$ $\beta$-function contains terms also involving the
gauge coupling $g^2$ at one loop. Hence, to one loop
$\beta^{abcd}_\lambda(a,\lambda)$ is given by
\begin{eqnarray}\label{betaabcd}
\beta^{abcd}_\lambda(a,\lambda) &=& \left[ \frac{1}{4} \left(
\lambda^{abpq} \lambda^{cpdq} + \lambda^{apbq} \lambda^{cdpq} +
\lambda^{apcq} \lambda^{bpdq}
+ \lambda^{apdq} \lambda^{bpcq} \right) \right. \nonumber \\
&& \left. -~ 12 C_A \lambda^{abcd} a ~+~ 8 C_A f^{abp} f^{cdp} a^2
~+~ 16 C_A f^{adp} f^{bcp} a^2 ~+~ 96 d_A^{abcd} a^2 \right]\;,
\end{eqnarray}
such that in $d$ dimensions
\begin{equation}\label{betaabcdbis}
    \mu\frac{\p}{\p\mu}\lambda^{abcd}=-2\varepsilon\lambda^{abcd}+\beta^{abcd}\,.
\end{equation}
Another useful check is the observation that $\beta^{abcd}$ enjoys
the same symmetry properties as the tensor $\lambda^{abcd}$,
summarized in (\ref{abcd}).

It is worth noticing that $\lambda^{abcd}$~$=$~$0$ is \emph{not} a
fixed point due to the extra $\lambda^{abcd}$-independent terms. Put
another way, if we had not included the $\lambda^{abcd}$-interaction
term in the original Lagrangian, then such a term would be generated
at one loop through quantum corrections, meaning that in this case
there would have been a breakdown of renormalizability. Further, we
note that with the presence of the extra couplings, the two loop
term of this $\beta$-function is actually scheme dependent.

Finally, we turn to the two loop renormalization of the localized
operator itself, or equivalently of the mass $m$. The operator can be read off from (\ref{actions2})
and is given by
\begin{equation}\label{nloa}
    \mathcal{O}=\left(B_{\mu\nu}^a-\oB_{\mu\nu}^a\right)F_{\mu\nu}^a\;.
\end{equation}
To do this we extend the one loop calculation, \cite{Capri:2005dy},
by again inserting this operator into a $A^a_\mu$-$B^b_{\nu\sigma}$
$2$-point function and deducing the appropriate renormalization
constant $Z_\mathcal{O}$, defined by
\begin{equation}\label{nlob}
    \mathcal{O}_o=Z_\mathcal{O}\mathcal{O}\;.
\end{equation}
One significant advantage of the massless field approach is that
there is no mixing of this dimension three operator into the various
lower dimension two operators, which was evident in the algebraic
renormalization analysis, and would complicate this aspect of the
two loop renormalization. In other words following the path of using
massive propagators would have required us to address this mixing
issue. Hence, from the $5$ one loop and $131$ two loop Feynman
diagrams, we find the $\MSbar$ renormalization constant
\begin{eqnarray}\label{zop}
Z_{\cal O} &=& 1 ~+~ \left[ \frac{2}{3} T_F \Nf - \frac{11}{6} C_A
\right]
\frac{a}{\varepsilon} \nonumber \\
&& +~ \left[ \left( \frac{121}{24} C_A^2 + \frac{2}{3} T_F^2 \Nf^2 -
\frac{11}{3} T_F \Nf C_A \right) \frac{a^2}{\varepsilon^2} ~+~
\left( \left( \frac{1}{3} T_F \Nf C_A - \frac{77}{48} C_A^2 + T_F
\NF C_F \right) a^2
\right. \right. \nonumber \\
&& \left. \left. ~~~~~+~ \frac{1}{512N_A} \lambda^{abcd}
\lambda^{acbd} - \frac{1}{32N_A} f^{abe} f^{cde} \lambda^{adbc} a
\right) \frac{1}{\varepsilon} \right]\;,
\end{eqnarray}
and therefore,
\begin{eqnarray}\label{resgammaop}
\gamma_{\cal O}(a,\lambda) =
&-&2\left(\frac{2}{3}T_FN_f-\frac{11}{6}C_A\right)a
    -\left(\frac{4}{3}T_FN_fC_A+4T_FN_fC_F-\frac{77}{12}C_A^2\right)a^2
    \nonumber\\&+&\frac{1}{8N_A}f^{abe}f^{cde}\lambda^{adbc}a-\frac{1}{128N_A}\lambda^{abcd}\lambda^{abcd}\;,
\end{eqnarray}
as the two loop $\MSbar$ anomalous dimension, which is defined as
\cite{Capri:2005dy}
\begin{equation}\label{rgeop}
\gamma_\mathcal{O}(a,\lambda)=\mu\frac{\p}{\p\mu}\ln
Z_\mathcal{O}\;.
\end{equation}
As at one loop it is independent of the gauge parameter, as expected
from the fact that the operator is gauge invariant. Also, the two
loop correction depends on the $\lambda^{abcd}$ couplings as well as
the gauge coupling, as expected from our earlier arguments. We end
this section by mentioning that a factor of $(-2)$ was erroneously
omitted in the one loop anomalous dimension $\gamma_\mathcal{O}(a)$
in eq.(6.9) of \cite{Capri:2005dy}.

\section{Equivalence between the massless theory and usual Yang-Mills theory.}
In this section, we shall discuss the usefulness of the
$\delta_s$-supersymmetry, defined by (\ref{ss}), and show that Green
functions which are built from the original Yang-Mills fields
\footnote{We shall not consider matter (spinor) fields in this
section, although the derived results remain valid.}
$\left\{A_\mu^a,c^a,\occ^a,b^a\right\}$ are independent from the
tensor coupling $\lambda^{abcd}$, in the massless version of the
physical model (\ref{completeaction}).

Next to this result, we shall also prove the stronger result that
any Green function constructed from the original Yang-Mills fields
$\left\{A_\mu^a,c^a,\occ^a,b^a\right\}$, evaluated with respect to
the massless version of our action, gives the \emph{same} result as
if the Green function would be evaluated with the original
Yang-Mills action.

\subsection{The massless case $m\equiv0$.}
As we have already noticed in (\ref{loc2bis}), the case corresponded
originally to the introduction of a unity into the usual Yang-Mills
action. Evidently, we expect that the model obtained with $m\equiv0$
would be exactly equivalent to ordinary Yang-Mills theory.
Nevertheless, this statement is a little less clear if we take a
look at the massless action
\begin{eqnarray}
% \nonumber to remove numbering (before each equation)
  S_{phys}^{m\equiv0} &=& S_{YM}+S_{gf}\nonumber\\&+&\int d^4x\left[\frac{1}{4}\left( \overline{B}_{\mu \nu
}^{a}D_{\sigma }^{ab}D_{\sigma }^{bc}B_{\mu \nu
}^{c}-\overline{G}_{\mu \nu }^{a}D_{\sigma }^{ab}D_{\sigma
}^{bc}G_{\mu \nu }^{c}\right)+
\frac{\lambda^{abcd}}{16}\left( \overline{B}_{\mu\nu}^{a}B_{\mu\nu}^{b}-\overline{G}_{\mu\nu}^{a}G_{\mu\nu}^{b}%
\right)\left( \overline{B}_{\rho\sigma}^{c}B_{\rho\sigma}^{d}-\overline{G}_{\rho\sigma}^{c}G_{\rho\sigma}^{d}%
\right) \right]\;,\label{dd1}
\end{eqnarray}
which is obtained from (\ref{completeaction}). The reader shall
notice that the quartic interaction $\propto\lambda^{abcd}$ is
anyhow generated, making the path integration over the tensor fields
no longer an exactly calculable Gaussian integral. Hence, we could
worry about the fact the tensor coupling $\lambda^{abcd}$ might
enter the expressions for the Yang-Mills Green functions, which are
those built out of the fields $A_\mu^a$, $c^a$, $\occ^a$ and $b^a$,
when the partition function corresponding to the action (\ref{dd1})
would be used.

We recall here that the action (\ref{dd1}) is invariant under the
supersymmetry (\ref{ss}). This has its consequences for the Green
functions. Let us explore this now. Firstly, we consider a generic
$n$-point function built up only from the original fields
$\left\{A_\mu^a,c^a,\occ^a,b^a\right\}$. More precisely, we set
\begin{eqnarray}\label{dd2}
    \mathcal{G}_n(x_1,\ldots,x_n)&=&\left\langle G_n(x_1,\ldots,x_n)\right\rangle_{S_{phys}^{m\equiv0}}\nonumber\\
    &=&\int D\Phi G_n(x_1,\ldots,x_n) e^{-S_{phys}^{m\equiv0}}\;,
\end{eqnarray}
with
\begin{eqnarray}\label{dd3}
G_n(x_1,\ldots,x_n)&=& A(x_1)\ldots
A(x_i)\occ(x_{i+1})\ldots\occ(x_{j})c(x_{j+1})\ldots
c(x_{k})b(x_{k+1})\ldots
    b(x_n)\;,
\end{eqnarray}
and we introduced the shorthand notation $\Phi$ denoting all the
fields. We notice that $\delta_s G_n=0$ whereas $G_n\neq
\delta_s\left(\ldots\right)$, i.e. any functional of the form
(\ref{dd3}) belongs to the $\delta_s$-cohomology.

We are interested in the dependence of $\mathcal{G}_n$ on
$\lambda^{abcd}$. A small computation leads to
\begin{eqnarray}\label{dd4}
    \frac{\p \mathcal{G}_n}{\p \lambda^{abcd}}&=&-\frac{1}{16}\int D\Phi G_n(x_1,\ldots,x_n)\int d^4x      \left( \overline{B}_{\mu \nu }^{a}B_{\mu \nu }^{b}-\overline{G}_{\mu \nu }^{a}G_{\mu \nu }^{b}%
\right)\left( \overline{B}_{\rho \sigma }^{c}B_{\rho \sigma}^{d}-\overline{G}_{\rho \sigma}^{c}G_{\rho \sigma}^{d}%
\right) e^{-S_{phys}^{m\equiv0}}\nonumber\\
&=& -\frac{1}{16}\int D\Phi G_n(x_1,\ldots,x_n)\delta_s \int
d^4x\left[
\left( \overline{B}_{\mu \nu }^{a}B_{\mu \nu }^{b}-\overline{G}_{\mu \nu }^{a}G_{\mu \nu }^{b}%
\right)\left(\overline{G}_{\rho\sigma}^{c}B_{\rho\sigma}^d\right)\right]e^{-S_{phys}^{m\equiv0}}\nonumber\\
&=&-\frac{1}{16}\int D\Phi \delta_s\left\{G_n(x_1,\ldots,x_n)\int
d^4x\left[
\left( \overline{B}_{\mu \nu }^{a}B_{\mu \nu }^{b}-\overline{G}_{\mu \nu }^{a}G_{\mu \nu }^{b}%
\right)\left(\overline{G}_{\rho\sigma}^{c}B_{\rho\sigma}^d\right)\right]\right\}e^{-S_{phys}^{m\equiv0}}\nonumber\\
&=&-\frac{1}{16}\left\langle\delta_s\left[G_n(x_1,\ldots,x_n)\int
d^4x
\left( \overline{B}_{\mu \nu }^{a}B_{\mu \nu }^{b}-\overline{G}_{\mu \nu }^{a}G_{\mu \nu }^{b}%
\right)\left(\overline{G}_{\rho\sigma}^{c}B_{\rho\sigma}^d\right)\right]\right\rangle_{S_{phys}^{m\equiv0}}\nonumber\\
&=&0\;.
\end{eqnarray}
The last line of (\ref{dd4}) is based on the fact that $\delta_s$
annihilates the vacuum as it generates a symmetry of the model.

We have thus shown that all original Yang-Mills Green functions will
be independent of the tensor coupling $\lambda^{abcd}$. These are of
course the most interesting Green functions, gauge variant (like the
gluon propagator) or invariant (the physically relevant Green
functions). The previous result does not mean that we can simply set
$\lambda^{abcd}\equiv0$ and completely forget about the quartic
interaction $\propto \left( \overline{B}_{\mu\nu}^{a}B_{\mu\nu}^{b}-\overline{G}_{\mu\nu}^{a}G_{\mu\nu}^{b}%
\right)\left( \overline{B}_{\rho\sigma}^{c}B_{\rho\sigma}^{d}-\overline{G}_{\rho\sigma}^{c}G_{\rho\sigma}^{d}%
\right)$. There is a slight complication, as quantum corrections
reintroduce the quartic interaction $\propto \lambda^{abcd}$ even
when we set $\lambda^{abcd}=0$ \footnote{This could also be
reinterpreted by stating that the bare coupling $\lambda_0^{abcd}$
is not proportional to the renormalized coupling $\lambda^{abcd}$.}.

\subsection{Exact equivalence between the massless action and the Yang-Mills action.}
In this subsection, we shall use once more the nilpotent $\delta_s$
symmetry to actually prove that
\begin{equation}\label{hom1}
    \left\langle G_n(x_1,\ldots,x_n)\right\rangle_{S_{YM}+S_{gf}}\equiv \left\langle
    G_n(x_1,\ldots,x_n)\right\rangle_{S_{phys}^{m\equiv0}}\;,
\end{equation}
meaning that the expectation value of any Yang-Mills Green function,
constructed from the fields $\left\{A_\mu^a,c^a,\occ^a,b^a\right\}$
and calculated with the original (gauge fixed) Yang-Mills action
$S_{YM}+S_{gf}$, is identical to the one calculated with the
massless action $S_{phys}^{m\equiv0}$, where it is tacitly assumed
that the gauge freedom of both actions has been fixed by the same
gauge fixing.

In this context, let us also mention that the physical content of
the massless theory described by $S_{phys}^{m\equiv0}$ will not
depend on the extra tensor fields, as $(B_{\mu\nu}^a,G_{\mu\nu}^a)$
and $(\overline{B}_{\mu\nu}^a,\overline{G}_{\mu\nu}^a)$ are both
$\delta_s$-doublets, and hence any physical operator shall certainly
not depend on these fields. Physical operators belong to the
$\delta_s$-cohomology, which is independent of $\delta_s$-doublets
\cite{Piguet:1995er}. More precisely, Yang-Mills theory and the
massless action (\ref{dd1}) will have identical physical operators,
which belong to the BRST $\widetilde{s}$-cohomology. Next to this,
we also know that the extra fields of the massless model shall
decouple from the physical spectrum as they belong to the trivial
part of the additional $\delta_s$-cohomology.

Let us now prove the statement (\ref{hom1}). We shall first prove
the following.
\begin{itemize}
\item[Theorem I] Let $S_0$ be an action constructed from a set
of fields $\phi_i$, that enjoys a symmetry generated by a nilpotent
operator $\delta$. Consider a second action $S_1=S_0+\Delta S$,
whereby $\Delta S$ is constructed from the fields $\phi_i$, and an
extra set $(\varphi_k,\overline{\varphi}_k)$ whereby $\varphi_k$ and
$\overline{\varphi}_k$ are $\delta$-doublets, such that $S_1$ also
enjoys the symmetry generated by $\delta$. We assume that the
renormalizability of $S_0$ and $S_1$ has been established.

The physical operators of $S_0$ and $S_1$ both belong to the
$\delta$-cohomology, whereas the extra fields
$(\varphi_k,\overline{\varphi}_k)$ do not. This is due to the fact
that these fields give rise to pairs of $\delta$-doublets, thus
having vanishing cohomology. The difference $\Delta S=S_1-S_0$ is
then also necessarily $\delta$-exact. Let $\mathcal{H}$ be an
operator belonging to the $\delta$-cohomology. Then we can write
\begin{equation}\label{hom2}
    \left\langle
    \mathcal{H}\right\rangle_{S_1}=\int d\phi_id\varphi_k\mathcal{H}e^{-S_0-\Delta S}\;.
\end{equation}
We define a new action
\begin{equation}\label{hom3}
    S_\kappa=S_0+\kappa\Delta S\;,
\end{equation}
where $\kappa$ is a global \footnote{It is tacitly assumed here that
$\Delta S$ is, disjoint from $S_0$, invariant with respect to other
possible symmetries. If not, we can evidently not introduce such a
global factor $\kappa$.} parameter put in front of the action-part
$\Delta S$. If we multiply the counterterm part, corresponding to
$\Delta S$, which belongs to a trivial part of the
$\delta$-cohomology, by $\kappa$, then the renormalizability will be
maintained, without the need of introducing a counterterm for
$\kappa$, so that $\kappa_0\equiv\kappa$. The parameter $\kappa$ can
be used to switch on/off the difference $\Delta S$. More precisely,
we can interpolate continuously between $S_0$ and $S_1$. Using this,
it is not difficult to show that
\begin{equation}\label{hom4}
    \frac{\partial}{\partial \kappa}\left\langle
    \mathcal{H}\right\rangle_{S_\kappa}=0\;,
\end{equation}
due to the $\delta$-exactness of $\Delta S$ and $\delta$-closedness
of $\mathcal{H}$. As a consequence
\begin{equation}\label{hom5}
    \left\langle
    \mathcal{H}\right\rangle_{S_0}=\left\langle
    \mathcal{H}\right\rangle_{S_1}\;.
\end{equation}
\end{itemize}
A related theorem is the following.
\begin{itemize}
\item[Theorem II] Let $S_0$ be an action constructed from a set
of fields $\phi_i$. Consider a second action $S_1=S_0+\Delta S$,
whereby $\Delta S$ is constructed from the fields $\phi_i$, and an
extra set $(\varphi_k,\overline{\varphi}_k)$ whereby $\varphi_k$ and
$\overline{\varphi}_k$ are $\delta$-doublets, such that $S_1$ enjoys
the symmetry generated by the nilpotent operator $\delta$. We
trivially extend the action of $\delta$ on the fields $\phi_i$ as
$\delta\phi_i=0$, so that of course $\delta S_0=0$. We assume that
the renormalizability of $S_0$ and $S_1$ has been established.

The difference $\Delta S=S_1-S_0$ is necessarily $\delta$-exact.
Moreover, the physical operators of $S_1$ must belong to the
$\delta$-cohomology, which is independent of the $\delta$-doublets
$(\varphi_k,\overline{\varphi}_k)$. The operators constructed from
the fields $\phi_i$, i.e. \emph{all} operators of the model $S_0$,
certainly belong to the $\delta$-cohomology of $S_1$.  Let
$\mathcal{K}$ be such an operator. A completely similar argument as
used in Theorem I allows to conclude that
\begin{equation}\label{hom2}
    \left\langle
    \mathcal{K}\right\rangle_{S_0}=\left\langle
    \mathcal{K}\right\rangle_{S_1}\;.
\end{equation}
\end{itemize}
Let us now comment on the usefulness of the previous theorems.
Theorem II is applicable to the Yang-Mills action $S_{YM}+S_{gf}$
and the massless action $S_{phys}^{m\equiv0}$, where $\delta_s$ is
the nilpotent symmetry generator of $S_{phys}^{m\equiv0}$, trivially
acting on $S_{YM}+S_{gf}$. Said otherwise, we have just proven that
the massless physical model (\ref{dd1}) is equivalent with
Yang-Mills, in the sense that the Green functions of the original
Yang-Mills theory remain unchanged when evaluated with the action
(\ref{dd1}).

An important corollary of the previous result is that the running of
the gauge coupling $g^2$ will be dictated by the usual
$\beta(g^2)$-function known from common literature, as the
renormalization factor for $g^2$ can be extracted from original
Yang-Mills $n$-point Green functions. This result was confirmed in
section III, as well as the fact that the other Yang-Mills
renormalization group functions remain unaltered, again in agreement
with Theorem II.

\subsection{The massive case $m\neq0$.}
It might be clear that the $\delta_s$-supersymmetry was the key tool
to prove the $\lambda^{abcd}$ independence in the massless case, as
well as the equivalence with Yang-Mills theory.

Evidently, we are more interested in the case that the mass $m$ is
present. The question arises what we may say in this case, as the
supersymmetry is now explicitly broken, i.e.
\begin{equation}\label{mc1}
    \delta_s S_{phys}\neq0\,,
\end{equation}
with the massive physical action given in (\ref{completeaction}).
The role of the tensor fields
$\left\{B_{\mu\nu}^a,\oB_{\mu\nu}^a,G_{\mu\nu}^a,\oG_{\mu\nu}^a\right\}$
remains an open question, as they do not longer constitute a pair of
doublets. But, we repeat, gauge or BRST invariance is kept, at both
the classical and quantum level.

\section{Conclusion and outlook.}
Returning to the rationale behind the construction of the action
(\ref{completeaction}), we recall that is was based on the
localization procedure, given in (\ref{loc2}), of the nonlocal
operator coupled to the Yang-Mills action, as displayed in
(\ref{ymop}). Clearly, we can return from the local physical action
to the original nonlocal one only if the extra couplings
$\lambda_1$, $\lambda_3$ and $\lambda^{abcd}$ would be zero. As
already explained, even if we set these couplings equal to zero from
the beginning, quantum corrections will reintroduce their
corresponding interactions.

Moreover, as the massive physical action is not
$\delta_s$-supersymmetric, we cannot simply say that we can choose
the extra couplings freely. Trying to get as close as possible to
the original localized version (\ref{actions2}) of the nonlocal
action (\ref{ymop}) where only one coupling constant is present, we
can imagine taking the tensor coupling $\lambda^{abcd}$ to be of the
form
\begin{equation}\label{dd8}
    \lambda^{abcd}=\ell_0^{abcd}g^2+\hbar\ell_1^{abcd}g^4+\ldots\,,
\end{equation}
i.e. we make it a series in the gauge coupling $g^2$, where the
coefficients $\ell_i^{abcd}$ are constant color tensors with the
appropriate symmetry properties as (\ref{jacobigen}) and
(\ref{abcd}). This already avoids the introduction of an independent
coupling. We have temporarily reintroduced the Planck constant
$\hbar$ to make clear the distinction between classical and quantum
effects. At the classical level, we could try to set
$\ell_0^{abcd}=0$ in order to kill the quartic interaction. Doing
so, we would at least keep the classical equivalence between the
local and nonlocal actions (\ref{completeaction}) and (\ref{ymop})
by employing the classical equations of motion. Unfortunately, this
is not possible. We should assure that (\ref{dd8}) is consistent
with the quantum model, meaning that we should assure the
consistency with the renormalization group equations. Taking the
quantum effects into account, next to the divergent contributions
canceled by the available counterterms, there will be also quantum
corrections to the quartic interaction which are finite but
nonvanishing in $\hbar g^2$. We can fix the classical (tree level)
value $\ell_0^{abcd}$ by demanding that the proposal (\ref{dd8}) is
consistent with the renormalization group function
(\ref{betaabcdbis}), and likewise for the higher order coefficients
$\ell_i^{abcd}$, i.e. we should solve
\begin{equation}\label{dd20}
    \mu\frac{\p}{\p\mu}\lambda^{abcd}(g^2)=\beta(g^2)\frac{\p}{\p
    g^2}\lambda^{abcd}(g^2)=\beta^{abcd}(g^2)\;,
\end{equation}
order by order, with $\lambda^{abcd}(g^2)$ defined as in
(\ref{dd8}). By using the renormalization group function
(\ref{betaabcd}) it is apparent that $\ell_0^{abcd}=0$ is not a
solution.

It is interesting to notice that the classical action should already
contain the classical quartic coupling
$\lambda^{abcd}_{cl}=\ell_0^{abcd}g^2$ in order to allow for a
consistent extension of the model at the quantum level. This
classical value is in fact dictated by one loop quantum effects, as
it is clear from the lowest order term of (\ref{dd20}).

A completely similar approach could be used for the mass couplings
$\lambda_1$ and $\lambda_3$. We could eliminate the extra couplings
in favour of the gauge coupling, without making any sacrifice with
respect to the renormalization group equations. Of course, this is a
nontrivial task, as it should be checked whether for instance
(\ref{dd20}) possesses meaningful solution(s), as the coefficient
tensors $\ell_i^{abcd}$ should be at least realvalued, whereas the
uniqueness of the solution might be not evident.

The aforementioned procedure is not new, as we became aware of works
like \cite{Zimmermann:1984sx,Oehme:1995xg} and references therein,
where the reduction of couplings and its use were already studied.

In this paper, the role of the extra couplings was not the primary
motivation. The main purpose of this paper was to establish a
further study of the gauge model (\ref{completeaction}) itself. We
have reported on a few properties. We mentioned the existence of a
nilpotent BRST symmetry of the physical action
(\ref{completeaction}) and we have commented on the fact that the
physical action can be embedded into a more general action that
possesses an interesting supersymmetry amongst the new fields. We
briefly returned to the renormalizability, in particular on the
algebraic renormalization of the extra tensor coupling
$\lambda^{abcd}$ where we drew an analogy with, for instance the
Higgs inspired Coleman-Weinberg model. The physical action itself
only enjoys the supersymmetry in the massless limit, in which case
we were able to prove the equivalence with ordinary Yang-Mills
theory by using an argument based on the cohomology of that
supersymmetry. We also presented explicit results concerning the
renormalization group functions of the model evaluated in the
$\MSbar$ scheme: the anomalous dimensions of the original Yang-Mills
fields and parameters were calculated to two loop order and turned
out to be identical to the ones calculated in massless Yang-Mills,
supplemented with the same gauge fixing, in agreement with the
general argument that both models are equivalent in the massless
limit. The one-loop anomalous dimension of the tensor coupling
$\lambda^{abcd}$ has also been evaluated, giving explicit evidence
that $\lambda^{abcd}=0$ is not a fixed point of the model, and
finally also the two loop anomalous dimensions of the auxiliary
tensor fields as well as of the gauge invariant operator
$\mathcal{O}=(B-\oB)F$, given in (\ref{nloa}), or equivalently of
the mass $m$, have been calculated.

An interesting issue to focus on in the future would be the possible
effects on the gluon Green functions arising from the massive
physical model (\ref{completeaction}), to find out whether mass
effects might occur in the gluon sector as, for instance, in the
gluon propagator.

Returning to the necessity of introducing extra couplings, since we
cannot keep the equivalence between actions (\ref{ymop}) and
(\ref{completeaction}) due to these couplings, the relation with the
nonlocal gauge invariant operator $F\frac{1}{D^2}F$ has become
obscured. However, this is not an unexpected feature, as this
operator is highly nonlocal, due to the presence of the inverse of
the covariant Laplacian.  Nevertheless, we believe that the final
model (\ref{completeaction}) is certainly relevant per se. To some
extent, it provides a new example of a renormalizable massive model
for Yang-Mills theories, which is gauge invariant at the classical
level and when quantized it enjoys a nilpotent BRST symmetry.

There are several other remaining questions concerning our model. At
the perturbative level for example, it could be investigated which
(asymptotic) states belong to a physical subspace of the model, and
in addition one should find out whether this physical subspace can
be endowed with a positive norm, which would imply unitarity.
Although the resolution of this topic is under study, it is worth
remarking that the nilpotency of the BRST operator might be useful
in this context.

Of course, at the nonperturbative level, not much can be said at the
current time. The model is still asymptotically free, implying that
at low energies nonperturbative effects, such as confinement, could
set in. One could search for indications of confinement, similarly
as it is done for usual Yang-Mills gauge theories. An example of
such an indication is the violation of the spectral positivity, see
e.g. \cite{Dudal:2005na,Alkofer:2003jj,Cucchieri:2004mf}. Proving
and understanding the possible confinement mechanism in our model is
probably as difficult as for usual Yang-Mills gauge theories.

Finally, it would be interesting to find out whether this model
might be generated dynamically. A possibility would be to start from
the massless version of the action (\ref{completeaction}), which was
written down in (\ref{dd1}). After all, it is equivalent with
massless Yang-Mills, and we can try to investigate whether a
nonperturbative dynamically generated term $m(\overline{B}-B)F$
might emerge, which in turn could have influence on the gluon Green
functions.

\section*{Acknowledgments.}
The Conselho Nacional de Desenvolvimento Cient\'{i}fico e
Tecnol\'{o}gico (CNPq-Brazil), the Faperj, Funda{\c{c}}{\~{a}}o de
Amparo {\`{a}} Pesquisa do Estado do Rio de Janeiro, the SR2-UERJ
and the Coordena{\c{c}}{\~{a}}o de Aperfei{\c{c}}oamento de Pessoal
de N{\'\i}vel Superior (CAPES) are gratefully acknowledged for
financial support. D.~Dudal is a postdoctoral fellow of the
\emph{Special Research Fund} of Ghent University. R.~F.~Sobreiro
would like to thank the warm hospitality at Ghent University, and
D.~Dudal that at the UERJ where parts of this work were prepared.

\end{document}